\documentclass[12pt,a4paper]{article}
\usepackage{amsmath,amsfonts,amssymb,amsthm,epsfig,epstopdf,titling,url,array}
\usepackage{geometry}
\usepackage{amsmath}
\usepackage{float}
\usepackage[english]{babel}
\usepackage{enumitem}
\usepackage{caption}
\usepackage{csquotes}
\usepackage[margin=8pt,font=small,labelfont=bf]{caption}
\usepackage[colorlinks,
            linkcolor=blue,
            urlcolor=blue,
            anchorcolor=blue,
            citecolor=blue
            ]{hyperref}
\usepackage{color}
\usepackage[normalem]{ulem}
\usepackage{algorithm}
\usepackage{algpseudocode}

\theoremstyle{plain}

\theoremstyle{remark}

\DeclareUnicodeCharacter{1D0C}{*************************************}
\title{Large Language Model in Financial Regulatory Interpretation}
\author{Zhiyu Cao\thanks{Stevens Institute of Technology, School of Business, Hoboken, NJ 07030, USA} 
\and 
Zachary Feinstein\thanks{Stevens Institute of Technology, School of Business, Hoboken, NJ 07030, USA, {\tt zfeinste@stevens.edu}}}

\date{\today}

\usepackage{graphicx}
\usepackage{subcaption}
\usepackage[toc,page]{appendix}

\usepackage[style=numeric]{biblatex}
\usepackage[nottoc]{tocbibind}
\DeclareFieldFormat[inbook]{citetitle}{#1}
\DeclareFieldFormat[inbook]{title}{#1}

\begin{document}

\maketitle

\begin{abstract}
This study explores the innovative use of Large Language Models (LLMs) as analytical tools for interpreting complex financial regulations. The primary objective is to design effective prompts that guide LLMs in distilling verbose and intricate regulatory texts, such as the Basel III capital requirement regulations, into a concise mathematical framework that can be subsequently translated into actionable code. This novel approach aims to streamline the implementation of regulatory mandates within the financial reporting and risk management systems of global banking institutions. A case study was conducted to assess the performance of various LLMs, demonstrating that GPT-4 outperforms other models in processing and collecting necessary information, as well as executing mathematical calculations. The case study utilized numerical simulations with asset holdings -- including fixed income, equities, currency pairs, and commodities -- to demonstrate how LLMs can effectively implement the Basel III capital adequacy requirements.
\end{abstract}
\clearpage

\section{Introduction}
Over the last decade, Artificial Intelligence (AI) and Natural Language Processing (NLP) models have been developed to process massive amounts of financial text, providing professional investment suggestions and aiding in financial decision-making. Research highlights that machine learning algorithms, including Long Short-Term Memory (LSTM) \cite{mikolov2013advances}, Convolutional Neural Networks (CNN) \cite{das2014text}, Support Vector Machines (SVM) \cite{gentzkow2019text}, Random Forest (RF) \cite{mashrur2020machine}, and the Bidirectional Encoder Representations from Transformers (BERT) \cite{achitouv2023natural, bochkay2023textual}, exhibit strong performance in handling financial texts. Moreover, several financial service firms also focus on developing AI tools to provide powerful support for participants in the financial markets by offering deep market insights and predictive analysis. A prime example is Bloomberg's AI platform, which analyzes data from global stock markets, news reports, and social media trends to predict the future trajectory of specific stocks or sectors.

Large Language Models (LLMs), epitomized by tools like ChatGPT, have marked a revolutionary shift in artificial intelligence. Recently, LLMs such as GPT-3.5 and GPT-4 have demonstrated a remarkable ability to comprehend and generate human-like text. FinBERT, an NLP tool adapted to the finance domain, has been shown to outperform other NLP models in identifying discussions related to environmental, social, and governance (ESG) texts \cite{huang2023finbert}. Meanwhile, a retrieval-augmented LLM framework for financial sentiment analysis has been introduced \cite{zhang2023enhancing}. The efficacy of LLM-based chatbots for personal finance advisement has also been assessed \cite{lakkaraju2023llms}. BloombergGPT, a model that leverages Bloomberg's vast domain-specific dataset, demonstrates that it outperforms existing models on general financial tasks \cite{wu2023bloomberggpt}. Given the impressive performance of LLMs, there is a natural impetus to explore the application of LLMs in financial regulatory interpretation.

Despite their groundbreaking advancements, Large Language Models (LLMs) also exhibit several limitations, primarily due to their nascent stage of development. First, preliminary models such as GPT-3.5, LLaMA-7B, and text-focused LLMs like Claude-3, are constrained by their inability to perform complex mathematical calculations or code analysis. Second, LLMs are highly sensitive to variations in prompts and document loading methods, which may result in significantly divergent outcomes for the same topic. If prompts are overly simplistic or lack precise direction, LLMs frequently generate inaccurate results. Third, the considerable number of parameters within these models makes pre-training and application both challenging and costly, necessitating extensive datasets and significant computational resources.

In response to these challenges, this study introduces the potential of LLMs in the interpretation of financial regulations. Our approach includes several key strategies:

First, we conduct a performance comparison among LLMs by manually examining a dataset to assess the capabilities of GPT-3.5, GPT-4, Gemini-1.5, and Claude-3. This assessment focuses on their ability to collect and analyze contextual information within financial regulatory documents. Through accuracy comparisons, we identify the most effective model. Second, our comparative analysis of document loading methods examines the efficacy of GPT-4 when analyzing financial regulatory documents uploaded as PDF files versus images. Our findings indicate that GPT-4 demonstrates greater precision in interpreting images than PDFs, particularly when documents contain a mix of mathematical equations, textual explanations, and footnotes, which typically present challenges for LLMs. Third, we develop an engineering method for prompt design. By incorporating key elements into prompts, we guide LLMs to analyze documents more accurately. Contrasting the performance of naive prompts with those crafted using our engineering method underscores the critical role of deliberate prompt design in the accuracy of information collection. Finally, through comprehensive case studies, we validate the application of LLMs in financial regulation. By employing proper prompt design and document loading methods, alongside selecting an appropriate LLM, we demonstrate accurate computation of the minimum capital requirements from the 'Minimum Capital Requirements for Market Risk' section of the Basel III framework.

The remainder of this paper is organized as follows: Section 2 provides an overview of related work concerning the application of LLMs in the finance field. Section 3 details a comprehensive framework for applying LLMs to financial regulation documents, including algorithms, prompt design, and document loading methods. Section 4 presents our dataset, comparative analysis across different LLMs, and a comprehensive case study. Section 5 examines the ethical considerations surrounding the application of LLMs in financial regulation, addressing critical aspects such as data privacy, transparency, and fairness. Finally, Section 6 concludes the paper by highlighting potential limitations and proposing avenues for future research.

\section{Literature Review: Large Language Models in Finance}\label{sec:literature}

The integration of LLMs into the financial sector represents a significant and emerging area of research. Numerous studies have focused on developing and applying domain-specific LLMs to enhance financial tasks such as sentiment analysis, textual analysis, and stock market predictions. Notably, FinBERT \cite{huang2023finbert}, which adapts the BERT framework specifically for the finance domain, has been shown to outperform other NLP models in identifying discussions related to financial texts. Additionally, PIXIU \cite{xie2023pixiu} proposes a comprehensive framework that represents the first financial LLM based on fine-tuning LLaMA with instructional data, enabling the model to execute various financial tasks effectively. InvestLM \cite{yang2023investlm} and FinGPT \cite{yang2023fingpt} have each contributed uniquely to areas such as market analytics and predictive accuracy. BloombergGPT \cite{wu2023bloomberggpt} leverages Bloomberg's vast domain-specific dataset, including news, market forecasts, and regulatory information. Furthermore, a retrieval-augmented LLM framework for financial sentiment analysis was introduced \cite{zhang2023enhancing}, and the efficacy of LLM-based chatbots for personal finance advisement was assessed \cite{lakkaraju2023llms}. A decision framework that aids financial professionals in selecting the most suitable LLM solutions based on their specific needs around data, computing power, and performance objectives was provided \cite{li2024flexkbqa}. Additionally, insights into Natural Language Processing techniques within the framework of financial regulation were offered \cite{achitouv2023natural}.

Despite the substantial focus on the development and application of domain-specific LLMs to enhance various financial tasks, their use in interpreting financial regulation documents remains relatively unexplored. This research addresses this gap and contributes to the existing literature in two distinct ways.

First, we investigate the innovative application of Large Language Models as analytical tools for interpreting complex financial regulations. By designing appropriate prompts and employing the correct document loading methods, we guide LLMs to distill verbose and intricate regulatory texts, such as the Basel III capital requirements, into a concise mathematical framework that is then translated into actionable code. Experimental results, discussed in subsequent sections, demonstrate the feasibility and accuracy of our method. This approach has the potential to streamline the implementation of regulatory mandates within the financial reporting and risk management systems of global banking institutions. Second, the Basel III standards necessitate advanced internal systems for risk assessment and management, substantial reporting obligations, and rigorous compliance protocols. These standards pose significant challenges for banks in maintaining higher operational and administrative capacities, particularly in identifying and interpreting regulatory requirements and integrating compliance workflows. Considering the limited resources available to small and medium-sized financial firms, our approach offers a potential solution by improving the identification of essential information and enhancing the efficiency of regulatory compliance processes.

\section{Framework}\label{sec:Framework}

Financial regulation documents are inherently complex, containing dense legal terminology, textual descriptions, mathematical formulations, and numerous footnotes. The high sensitivity of large language models presents challenges in achieving perfect accuracy in information retrieval from such documents.

This section describes a systematic algorithm designed for the analysis of financial regulation documents. The process begins with the efficient loading of relevant documents, followed by prompt engineering to define the overall problem-solving process. For complex financial tasks, achieving accuracy typically requires multiple iterative steps. Each task is broken down into smaller, manageable objectives, with each being addressed through carefully crafted prompts designed to secure accurate outcomes.

Although LLMs provide preliminary insights and help identify specific data locations, initial results may not always be accurate. We manually verify all key information. If the retrieved information is incorrect or unavailable, we activate an additional mechanism to locate accurate data sources. Following this, we manually upload relevant documents, aiding the LLMs in refining their analysis. Finally, we use the LLM for mathematical calculations to determine the correct outputs.

\begin{figure}[htbp]
\centerline{\includegraphics[width=\columnwidth]{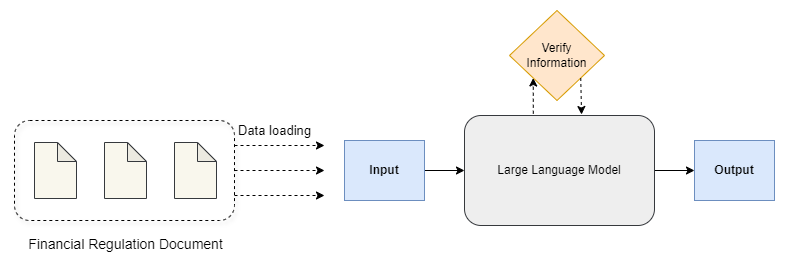}}
\caption{Visual representation of the process for interpreting financial regulation documents.}
\label{fig:1}
\end{figure}

\begin{figure}[htbp]
\centerline{\includegraphics[width=\columnwidth]{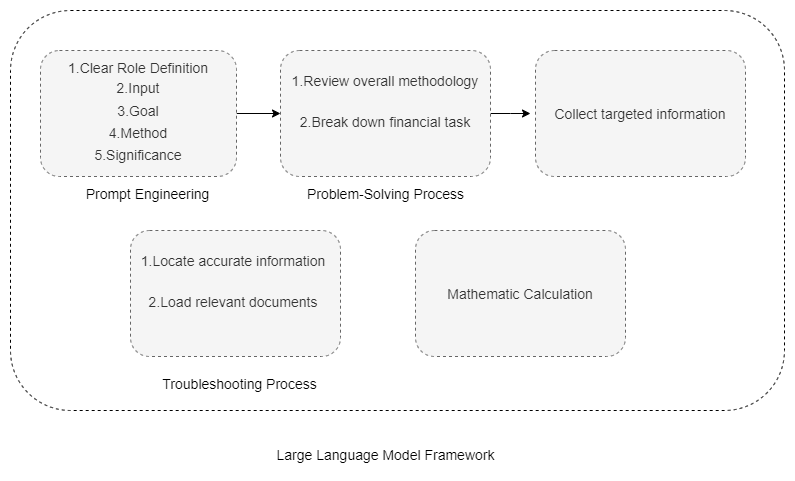}}
\caption{Detailed schematic of the Large Language Model used for financial document analysis.}
\label{fig:2}
\end{figure}

Figures~\ref{fig:1} and~\ref{fig:2} present a high-level illustration of the proposed architecture. We will elucidate prompt engineering, document loading methods, and the detailed process in the rest of this section.

\subsection{Document Loading Methods}\label{loading}

Initially, we utilized the plug-in feature within GPT-4 to process PDF files from the Basel III framework. While this method proved adequate for simple tasks, it faltered when handling more complex content, resulting in errors in LLM outputs.

Recognizing the limitations inherent to the PDF format—which primarily prioritizes presentation over content extraction—we sought alternative document loading techniques. Our exploration led to a significant finding: converting PDFs to images before uploading markedly improves GPT-4's analytical capabilities. Unlike PDFs, images circumvent the complexities tied to parsing diverse formats and layouts. This simplification allows the LLM to focus more on visual data extraction, which is inherently less prone to the errors typically introduced by the varied and intricate structures of PDFs. By adopting an image-based analysis approach, GPT-4 achieves a more reliable interpretation of mathematical formulas, charts, and tables, which often present challenges in PDFs due to their layered configurations.

In Section IV, we will compare the efficacy of PDF and image loading methods in enabling LLMs to accurately identify correlations between sensitivities.

\subsection{Prompt Design}

Research by \cite{du2023guiding, marvin2023prompt, giray2023prompt} highlights that prompt engineering is a relatively new discipline that focuses on developing and optimizing prompts to effectively utilize Large Language Models (LLMs), particularly in natural language processing tasks. This practice has emerged as an essential skill for effective communication and interaction with LLMs. The design of prompts significantly influences the performance of these models. Our study articulates key principles of prompt design to guide LLMs in the interpretation of financial regulation documents.

First, the prompt must incorporate the following key elements:

\begin{enumerate}
\item \textbf{Clear Role Definition}: Provides specific instructions that guide the model's behavior and outline a macro-level approach to problem-solving. For us, the role is for an LLM to act as a specialized financial regulations interpreter.
\item \textbf{Input}: Describes the document or question that we want the model to process and provide a response for. For us, the input is the Minimum Capital Requirements for Market Risk'' from Basel III issued by the Basel Committee on Banking Supervision.     \item \textbf{Goal}: Specifies the details of the desired output. For us, the goal is to transform complex financial regulations into clear mathematical representations while maintaining the integrity of the regulation's core principles.     \item \textbf{Method}: Outlines a general strategy for addressing the problem, wherein the model reads the document and enriches its response based on the contained information. For us, the method is to understand legal terminology, offer straightforward explanations, and summarize the key elements to complete tasks.     \item \textbf{Significance}: Highlights the overarching concept and importance of the task, analogous to providing tips" within the prompt to guide effective outcomes. For us, the significance is to bridge the gap between dense regulatory texts and practitioners with no legal background.
\end{enumerate}

In our prompt design, it is crucial to avoid ambiguity, bias reinforcement, overfitting, lack of context, and unrealistic dependency on model limitations \cite{giray2023prompt}. These considerations ensure the effectiveness of our prompts in applications related to financial tasks.

In Section IV, we will compare the efficacy of naive prompts with detailed prompts developed using our method to enable LLMs to accurately identify risk buckets, risk weights, and correlations.

\section{Case Studies}\label{sec:cs}

\subsection{Dataset}\label{dataset}
In accordance with the Basel III framework, we conducted simulations on over 40 different asset holdings to evaluate the capability of LLMs in interpreting financial regulation documents. The document used for this analysis includes a comprehensive section on `Minimum Capital Requirements for Market Risk',\footnote{Bank for International Settlements, \textit{Minimum Capital Requirements for Market Risk}, \url{https://www.bis.org/bcbs/publ/d352.pdf}} which outlines the required capital reserves that banks must maintain to mitigate potential losses from market fluctuations. The document comprises approximately 184,000 tokens and includes legal terms, intricate calculations, and case analyses pertinent to diverse market risks, such as interest rate, equity, foreign exchange (FX), and commodity risks. To illustrate our methodology, Table~\ref{tab:asset-holdings} presents a typical simulated case, outlining a variety of bank asset holdings, ranging from treasury bonds and futures contracts to equities and currency pairs.

\begin{table}[ht]
    \caption{Bank Asset Holdings}
    \centering
    \begin{tabular}{|l|l|r|}
      \hline
      \textbf{Asset Type} & \textbf{Description} & \textbf{Quantity / Value} \\
      \hline
      U.S. Treasury Bond & 5-year & \$10,000 \\
      U.S. Treasury Bond & 10-year & \$10,000 \\
      Futures Contracts & Gold & 600 ounces \\
      Futures Contracts & Crude Oil & 2,000 barrels \\
      Equity & Exxon Mobil & 10,000 shares \\
      Equity & AT\&T & 10,000 shares \\
      Currency Pair & Long EUR/USD & 100,000 EUR \\
      Currency Pair & Short USD/JPY & 10,000,000 JPY \\
      \hline
    \end{tabular}
    \label{tab:asset-holdings}
\end{table}

\subsection{Minimal Capital Requirement Calculation}
In this case study, we provide a comprehensive analysis of the application of LLMs as analytical tools for interpreting the "Minimal Capital Requirement" section of the Basel III framework. This includes a detailed demonstration of how minimal capital requirements are calculated based on bank asset holdings, as illustrated in Table~\ref{tab:asset-holdings}.

We begin by uploading images of the relevant sections of the Basel III document, focusing on minimal capital requirements, into GPT-4. A custom-designed prompt is used to guide GPT-4 in extracting methodologies for calculating these requirements. The response from GPT-4 breaks down the task of calculating the Minimal Capital Requirement into the following manageable objectives:

\begin{enumerate}
    \item \textbf{Risk Classification:} Identify and categorize various types of risks associated with asset holdings, including equity risk, FX risk, general interest rate risk, and commodity risk.
    
    \item \textbf{Sensitivity Calculation:} Determine the sensitivity of each asset holding to different risk factors, such as changes in market prices, interest rates, and foreign exchange rates.
    
    \item \textbf{Aggregation of Risk Positions:} Combine individual risk positions to compile an overall risk profile.
    
    \item \textbf{Capital Requirement Calculation:} Compute the minimal capital requirement based on the aggregated risk profile, applying the specific risk weights and methodologies prescribed by the Basel III framework.
\end{enumerate}

In this report, we particularly focus on the Delta Equity Risk calculation. The same method is applied to other sections, repeating the process as necessary. Regarding the Sensitivity Calculation, through a carefully designed prompt, GPT-4 is capable of identifying relevant sources based on the context, specifically locating information regarding Delta Equity Sensitivities in Paragraph 67. We summarize GPT-4's findings as follows:

\textbf{Delta Sensitivity Calculation:} For an equity $k$, the sensitivity $s_k$ is defined as the change in the market value of instrument $i$ ($V_i$) with respect to a 1 \% change in the market value of equity $k$ ($EQ_k$). Mathematically, it is expressed as:
\[
s_k = \frac{V_i(1.01 \cdot EQ_k) - V_i(EQ_k)}{0.01}
\]
where:
\begin{itemize}
  \item $k$ is a given equity.
  \item $EQ_k$ is the market value of equity $k$.
  \item $V_i$ is the market value of instrument $i$ as a function of the price of equity $k$.
\end{itemize}

In this case study, GPT-4 can calculate the sensitivity of each equity:
\begin{itemize}
  \item Delta Sensitivity for ``Exxon Mobil'': \$1,100,000.
  \item Delta Sensitivity for ``AT\&T'': \$170,000.
\end{itemize}

In the Aggregation of Risk Positions step, GPT-4 is utilized to collect and analyze targeted information firstly.

\begin{itemize}
    \item \textbf{Exxon Mobil} is classified under the category \textit{Energy, Oil and Gas} within an advanced economy, and is placed in \textit{Bucket 7}.
    \item \textbf{AT\&T} is classified under \textit{Telecommunications} within an advanced economy, and is placed in \textit{Bucket 6}.
\end{itemize}

\textbf{Risk Weights and Correlations:}
\begin{itemize}
    \item The risk weight assigned to Bucket 7 (Exxon Mobil) is $40\%$.
    \item The risk weight assigned to Bucket 6 (AT\&T) is $35\%$.
    \item The prescribed correlation between Exxon Mobil and AT\&T is set at $15\%$.
\end{itemize}

This outcome is the result of GPT-4’s comprehensive capabilities, which include the identification of each equity and a thorough understanding of the Basel III framework. GPT-4 efficiently processes tables, textual descriptions, and mathematical formulas to produce these results.

GPT-4 successfully located information regarding Delta Equity risk calculation in Paragraph 51. Here is the response:
\subsection*{Calculation Within Delta Buckets}
\begin{itemize}
    \item \textbf{Weighted Sensitivity \((WS_k)\):}
    \[
    WS_k = RW_k \cdot S_k
    \]
    where \(RW_k\) is the risk weight for equity \(k\) and \(S_k\) is the sensitivity of equity \(k\).

    \item \textbf{Risk Position \((K_b)\) for Delta Bucket \(b\):}
    \[
    K_b = \sqrt{\sum_k WS_k^2 + \sum_k \sum_{k \neq l} \rho_{kl} WS_k WS_l}
    \]
    where \(\rho_{kl}\) represents the correlation between equities \(k\) and \(l\) within the same bucket.
\end{itemize}

\subsection*{Aggregating Delta Risk Positions Across Buckets}
\begin{itemize}
    \item \textbf{Delta Risk Charge Calculation:}
    \[
    \text{Delta} = \sqrt{\sum_b K_b^2 + \sum_b \sum_{c \neq b} \gamma_{bc} S_b S_c}
    \]
    where \(S_b = \sum_k WS_k\) for all risk factors in bucket \(b\), and \(S_c = \sum_k WS_k\) in bucket \(c\), and \(\gamma_{bc}\) is the correlation between buckets \(b\) and \(c\).
\end{itemize}

The calculated delta equity risk capital requirement, as shown in Table~\ref{tab:asset-holdings}, is approximately \$461,262.67. This result matches the outcome from our manually simulated dataset.

\subsection{Comparative Analysis across Different LLMs}\label{C}
In this case study, we investigate the capabilities of large language models (LLMs) in interpreting and applying complex financial regulations, based on the `Minimum Capital Requirements for Market Risk' document from Basel III. To conduct a comprehensive assessment, we divided the document into several sections based on its table of contents, focusing on general interest rate risk, equity risk, foreign exchange (FX) risk, and commodity risks. For each sector, we designed detailed prompts to guide the LLMs in identifying key elements necessary for calculating the Minimum Capital Requirements from provided asset holding cases, aiming to assess the models' ability to identify buckets, risk weights, and correlations. We evaluated the performance of four prominent LLMs: GPT-4, GPT-3.5, Claude-3, and Gemini-1.5-pro, on our manually simulated dataset including 40 different asset holdings. Their accuracy was measured by the number of cases correctly identified, scaled by the total number of cases in the testing sample. The results of this evaluation are presented in Table~\ref{table:comparison}.

\begin{table}[htbp]
\caption{Comparison of LLMs in Identifying Key Elements}
\label{table:comparison}
\centering
\scriptsize
\begin{tabular}{|l|c|c|c|}
\hline
\textbf{Model} & \textbf{Buckets (\%)} & \textbf{Risk Weights (\%)} & \textbf{Correlation (\%)} \\
\hline
GPT-4 & 85 & 100 & 96.5 \\
\hline
GPT-3.5 & 10 & 30 & 0 \\
\hline
Claude-3-Opus & 82.5 & 100 & 97.5 \\
\hline
Gemini-1.5-Pro & 27.5 & 75 & 80 \\
\hline
\end{tabular}
\end{table}

As indicated in Table~\ref{table:comparison}, GPT-4 and Claude-3-Opus achieve the highest overall performance among the models evaluated. Both models demonstrate near-perfect accuracies in identifying risk weights and correlations, with scores of 100\% and 96.5\% for GPT-4, and 100\% and 97.5\% for Claude-3-Opus, respectively. Additionally, GPT-4 slightly outperforms Claude-3-Opus in identifying buckets, with an accuracy of 85\% compared to 82.5\%. This suggests that these two models have a strong capability to comprehend and apply the complex rules and guidelines outlined in the `Minimum Capital Requirements for Market Risk' document. In contrast, GPT-3.5 and Gemini-1.5-Pro show lower performance across all three aspects. GPT-3.5 struggles to identify buckets (10\% accuracy) and fails to identify any correlations (0\% accuracy). Gemini-1.5-Pro, on the other hand, also struggles with identifying buckets (27.5\% accuracy) but performs moderately well in identifying risk weights (75\%) and correlations (80\%).

Recent research by \cite{yuan2023scaling, imani2023mathprompter, ahn2024large} suggests that mathematical reasoning poses significant challenges for large language models. In this work, we evaluated the mathematical computation capabilities of GPT-4, GPT-3.5, Claude-3, and Gemini-1.5-pro within the context of complex calculations related to the Minimum Capital Requirements (MCR) sector. We designed five distinct scenarios corresponding to key risk categories outlined in the regulatory document: general interest rate risk, equity risk, foreign exchange (FX) risk, and commodity risks.

\begin{table}[ht]
\caption{Accuracy of LLMs in Complex Mathematical Calculations}
\centering
\begin{tabular}{|l|c|}
\hline
\textbf{Model} & \textbf{Accuracy in MCR Calculations (\%)} \\ \hline
GPT-4 & 95 \\ \hline
GPT-3.5 & 0 \\ \hline
Claude-3-Opus & 38 \\ \hline
Gemini-1.5-Pro & 58 \\ \hline
\end{tabular}
\label{table:math_comparisons}
\end{table}

Table~\ref{table:math_comparisons} reveals a notable disparity in the accuracy of large language models when performing complex mathematical calculations for Minimum Capital Requirements (MCR). GPT-4 stands out as the most capable model, achieving an impressive accuracy score of 95\% across all tested scenarios. In contrast, GPT-3.5, while capable of providing a mathematical framework for the calculations, fails to execute the required complex operations, resulting in an accuracy score of 0\%. Claude-3-Opus and Gemini-1.5-Pro demonstrated moderate performance, with accuracy scores of 38\% and 58\%, respectively. Despite their ability to locate relevant information within the document, these models show significant room for improvement in performing complex mathematical calculations.

\subsection{Document Loading Method}\label{Case}
We selected Claude-3-Opus and GPT-4, which have performed well in identifying and interpreting regulatory requirements, to compare the performance of PDF loading and image loading. We conducted tests on these models using our manually simulated dataset to evaluate their ability to discern correlations between sensitivities. This task is inherently complex, as it typically involves intricate elements of Basel III such as mathematical formulas, legal terms, tables, and footnote analysis.

\begin{table}[ht]
\caption{Accuracy of Identifying Correlation from Different Document Loading Methods in Claude-3-Opus and GPT-4}
\label{table:accuracy_comparison}
\centering
\begin{tabular}{|l|c|c|}
\hline
\textbf{Model / Document Type} & \textbf{PDF (\%)} & \textbf{IMAGE (\%)} \\ \hline
Claude-3-Opus                  & 76.5              & 97.5                \\ \hline
GPT-4                          & 68                & 96.5                \\ \hline
\end{tabular}
\end{table}

Table~\ref{table:accuracy_comparison} displays the accuracy of Claude-3-Opus and GPT-4 in identifying correlations from documents loaded as PDFs and images. For documents loaded as PDFs, the accuracies recorded were 76.5\% for Claude-3-Opus and 68\% for GPT-4. Remarkably, when analyzing documents loaded as images, both models achieved impressive accuracies of 97.5\% and 96.5\%, respectively. Our troubleshooting process, outlined in Section III, revealed issues when PDF-loaded documents included mathematical formulas, legal terms, tables, and footnotes. This observation suggests that image-based document loading may be particularly effective for LLMs. This could be potentially due to the visual processing capabilities inherent in their architectures, which might better handle layouts and embedded information such as mathematical formulas that are typically difficult to parse in text-based PDF files.

\subsection{Naive Prompt vs Detailed Prompt}\label{subsec:prompt-comparison}
We compare the efficacy of naive prompts with detailed prompts developed using our method, in enabling LLMs to accurately identify risk buckets, weights, and correlations in the GPT-4 Model.

\begin{table}[ht]
\caption{Comparison of Naive Prompt and Detailed Prompt in Identifying Key Elements with GPT-4}
\label{table:1comparison}
\centering
\scriptsize 
\begin{tabular}{|l|c|c|c|}
\hline
\textbf{Model} & \textbf{Buckets (\%)} & \textbf{Risk Weights (\%)} & \textbf{Correlation (\%)} \\
\hline
Naive Prompt & 65.5 & 100 & 30 \\
\hline
Our Detailed Prompt & 85 & 100 & 96.5 \\
\hline
\end{tabular}
\end{table}

Table~\ref{table:1comparison} illustrates the effectiveness of naive and detailed prompts in the GPT-4 model. While the naive prompt achieves high accuracy (100\%) in simpler tasks such as identifying risk weights, it exhibits considerable limitations in more complex tasks. For example, it only achieves a 65.5\% accuracy in bucket identification and 30\% accuracy in correlation determination. In contrast, the detailed prompt significantly enhances performance across more complex areas, maintaining 100\% accuracy in risk weights and notably improving to 85\% and 96.5\% in identifying buckets and correlations, respectively. This data underscores the importance of using detailed prompts when addressing tasks of higher complexity to avoid errors.

\section{LLM Ethical Considerations}
The integration of Large Language Models (LLMs) into financial regulatory frameworks necessitates a comprehensive examination of multifaceted ethical dimensions \cite{jiao2024navigating}, with particular emphasis on data privacy, transparency, and fairness. These critical considerations are expected to become central to the discourse on deploying LLMs within the financial regulatory sphere in the near future.

Firstly, data privacy emerges as a paramount concern, considering the highly sensitive nature of financial information. LLMs trained on financial regulation documents must strictly adhere to robust privacy protection measures, such as differential privacy techniques, to safeguard the confidential information of both individuals and institutions. For instance, the data used to compute market risks for financial institutions often encompasses proprietary information regarding their asset and liability levels. Implementing these rigorous privacy protocols can effectively prevent data breaches that could erode public trust in financial institutions, thereby undermining their credibility and potentially catalyzing widespread economic repercussions.

Secondly, transparency in the application of LLMs within the domain of financial regulation is a critical ethical consideration that warrants careful attention. To maintain public trust and ensure accountability, the development and deployment of LLMs in this context must be characterized by a high degree of transparency. This necessitates clear and comprehensive documentation of the data origins, training methodologies, and decision-making processes employed by these models. Regular publication of evaluation results and performance metrics is essential to facilitate external scrutiny and validate the integrity of LLM-driven financial regulatory systems. 

Thirdly, ensuring fairness in LLM-driven financial decision-making is paramount to prevent discriminatory outcomes based on factors such as the geographic location and size of financial institutions. Bias mitigation strategies, including dataset enhancement and adversarial learning, should be employed to promote equitable treatment of all financial institutions, regardless of their regional context or operational scale, in regulatory processes. LLMs must be designed to account for the unique challenges faced by smaller, regional financial institutions, ensuring that they are not unfairly disadvantaged compared to larger, multinational corporations. This may involve incorporating diverse datasets that adequately represent the full spectrum of financial institutions, from local credit unions to global banking conglomerates. Moreover, adversarial learning techniques can help identify and mitigate potential biases that could skew regulatory decisions in favor of certain types of institutions.

\section{Conclusion}
In this study, we have demonstrated the innovative application of Large Language Models such as GPT-4, Gemini-1.5-pro and Claude 3 in interpreting complex financial regulation documents. While achieving 100\% accuracy in extracting information from texts that include legal terms, textual descriptions, mathematical formulations, and extensive footnotes remains a formidable challenge, we have developed a troubleshooting process and we deconstruct a complex financial task into smaller, manageable objectives. This approach has enabled us to achieve precise results through detailed prompting. Additionally, we conducted a comparative analysis of document loading methods and the performance of LLMs in interpreting financial documents. We have also devised a systematic approach to prompt design, which enhances the LLMs' capability to analyze and summarize financial texts effectively.

Our work further highlights a pair of promising avenues for future research. Although our framework effectively performs well on our manually simulated dataset, our investigation focused on scenarios involving a limited number of cases. Real-world cases, however, may encompass a more extensive range of assets holding. Thus, future research can integrate more comprehensive, interconnected datasets to derive profound insights. Furthermore, incorporating stress testing by generating synthetic datasets and designing stress tests using LLMs could further enhance the robustness and regulatory applicability of our framework.

\newpage
\printbibliography
\end{document}